\documentclass[granma]{svjour}

\usepackage[latin1]{inputenc}
\usepackage{graphicx}

\newcommand{\eq}[1]{(\ref{eq:#1})}
\newcommand{\mt}{\mu_\mathrm{t}}
\newcommand{\mn}{\mu_\mathrm{n}}
\newcommand{\vt}{v_\mathrm{t}}
\newcommand{\Ft}{F_\mathrm{t}}
\newcommand{\Ftmax}{F_\mathrm{t,max}}
\newcommand{\Fn}{F_\mathrm{n}}

\newcommand{\Fc}{F_\mathrm{coh}}
\newcommand{\reff}{r_\mathrm{eff}}

\newcommand{\Ttmax}{T_\mathrm{t,max}}
\newcommand{\Tnmax}{T_\mathrm{n,max}}
\newcommand{\fig}[1]{Fig. \ref{fig:#1}}
\newcommand{\por}{E}

\begin{document}
\title{The effect of contact torques on porosity of cohesive powders}
\author{Guido Bartels\inst{1}\mail{bartels@comphys.uni-duisburg.de},
  Tam\'as Unger\inst{1,2}, Dirk Kadau\inst{1}, Dietrich  E.~Wolf\inst{1} and J\'anos
  Kert\'esz\inst{2} 
  \thanks{ This work was supported by the German Science Foundation (DFG)
  within SFB 445 ``Nano-Particles from the Gas Phase'', by the BMBF
  through grant HUN 02/011 and by Federal Mogul GmbH. We thank Z\'eno
  Farkas and Lothar Brendel for useful discussions.} }
\institute{Institute of Physics \\ University
  Duisburg-Essen \\ 47048~Duisburg, Germany \\ \\ Institute of Physics \\
  Technical University of Budapest\\ 1111 Budapest, Hungary \\ \\ PACS number(s): 45.70.Cc, 62.25.+g, 83.10.Rs }
\date{Received: \today / Revised version: date}

\maketitle

\begin{abstract}
The porosity of uniaxially compacted cohesive powders depends on the
applied stress (including gravity). The case, where these stresses are
weak, is considered. The compaction results in a porosity which is a
function of sliding, rolling and torsion friction. By contact dynamics
simulations it is shown that the influences of contact torques (static
rolling and torsion friction) on the porosity are significant 
and approximately additive. The relevance for nano-powder pressure
sintering is discussed. 

\keywords{Granular compaction, Cohesive powders, Contact dynamics
  simulations, Contact torques}
\end{abstract}

\section{Introduction}

The behavior of granular packings under external load
is governed by particle properties (e.g. roughness, elasticity,
cohesion) as well as by the 
geometrical structure of the packing (e.g. connectivity,
orientations of contacts).
When compacting loose granular material, density and connectivity 
increase until a static state is reached, where the material
withstands the external pressure. Such jammed states are currently
widely studied \cite{Cates98,Silbert02,Makse04}.  

For noncohesive materials the porosity in such a jammed state depends
essentially on the deformability of the grains under the fixed
external load. If the particles were rigid, the porosity could not be
reduced by upscaling the external load (including gravity), but only
by shaking \cite{Knight95,Caglioti97}.  

However, for cohesive powders, static states with much
higher porosity are possible. They are due to stabilization mechanisms, 
which involve the cohesion force as an intrinsic force
scale. Therefore, these states are {\em not} stable for arbitrary
upscaling of the external load, but only up to a threshold depending
on the porosity. Exceeding this threshold leads to further compaction
of the powder. In order to avoid confusion we call these static states
``blocked'' instead of jammed.

This paper considers stabilization mechanisms of blocked states in
three dimensional cohesive powders of rigid particles.
The relative motion of two solid spheres has six degrees of freedom, three
translational characterized by a velocity vector $\vec{v}$ with one normal
component  and two tangential components (sliding modes),
and three rotational characterized by an angular velocity vector
$\vec{\omega}$, again with one normal component (torsion mode) and two
tangential components (rolling modes). If any of these modes are damped or
blocked they represent specific dissipation or stabilization mechanisms,
respectively. Well
known examples are the static and sliding friction: the former stabilizes a
contact against small tangential forces (blocking), the latter dissipates
kinetic energy (damping) in case of sliding.

All modes of relative motion can be blocked: In addition to static
friction also resistance 
against separation, rolling and torsion must be taken into account, i.e.
the contact can exert also cohesion force, normal and tangential
torques (i.e. torsion and rolling torques, respectively) in
order to inhibit relative motion. To what extent these particle interactions
stabilize pores in a cohesive powder is the
subject of this paper. 

Extending previous work in two dimensions \cite{Kadau2003}, we
introduce torsion 
friction in addition to rolling friction and cohesion into contact
dynamics simulations of a uniaxial compression process.  Of course,
two perfectly rigid spheres, if they existed, would only have a
contact {\em point}, which could neither exert a rolling nor a
torsion torque on the particles in contact. Even the Coulomb-Amonton-daVinci
friction law would not be justified for such an idealized point
contact. By contrast, we consider rigid spheres here only as a
geometrical idealization of real particles and do not take their
contact areas into account explicitly. Implicitly, however, the
finite size of the contacts is responsible for the various kinds of
friction we consider. 

It is believed that torsion and rolling resistance are of little importance
in noncohesive granular assemblies of spheres, where indeed in statics the
contact-torques vanish \cite{Brilliantov98} or are very weak
\cite{farkas2003}.  Our work was motivated rather by powders with grain
size smaller than $100$nm. The field of nano-powders \cite{hahn2003}
attracts a lot of scientific and industrial interest due to their different
material properties compared to assemblies of larger grains or bulk
materials. Many concepts developed for regular granular media can be
applied also to nano-powders, but additional aspects have to be taken into
account as well, such as strong cohesion (due to van der Waals forces)
and sinter-neck formation\cite{groza99} between the grains which make the
question of blocking torsion and rolling modes relevant. Little is known so far
about the way in which a sinter-neck resists rolling or
torsion. It is plausible, however, that on a sufficiently short time
scale (where creep can be neglected) torque thresholds have to be
exceeded to break a sinter-neck and induce relative motion between the
particles. Therefore, lacking well proved contact
laws on the nano-scale, we assume the simplest kind of threshold
dynamics for sliding, rolling and torsion friction.

\section{Model}
We consider rigid spherical particles of identical sizes, but, as
mentioned above, \emph{force} and also
\emph{torque} transmission is allowed at contacts. The total normal force
$F_{\rm n}$ 
between two grains has two constituents: first the attractive part, which
is a constant cohesion force $F_{\rm coh}$ and second the force $F_{\rm
  exc}$ due to the excluded volume constraint:
\begin{equation}
 F_{\rm n} = F_{\rm exc} - F_{\rm coh}\, ,
  \label{eq:Fn}
\end{equation} 
where the repulsion (attraction) is denoted by positive (negative)
sign. Depending on external forces acting on the particles the constraint
force $F_{\rm exc}$ can take any positive value. As a consequence the total
normal force can be repulsive $F_{\rm n} > 0$ (arbitrarily strong) or
attractive $0 > F_{\rm n} 
> - F_{\rm coh}$ as well and is in this latter case limited by the cohesion
force. Thus $F_{\rm n}$ inhibits detachment as long as the pulling is
weaker than the 
cohesion force $F_{\rm coh}$. If $F_{\rm coh}$ is exceeded, the contact starts to
open, but breaks only when the work done by the pulling force exceeds
the cohesion energy ($F_{\rm coh}$ times cohesion range).
More details of this implementation of
cohesion can be found in \cite{Kadau2002}. 

The tangential force, as well as the normal and tangential torques are
responsible for blocking the sliding, torsion and rolling modes,
respectively. Their implementation is described next.

\begin{figure}
   \centering
   \includegraphics[scale=0.4]{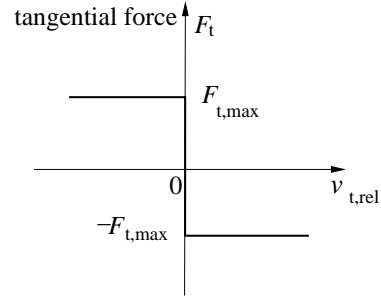}
   \caption{The graph represents Coulomb's law of friction: A sticking
    contact (with relative tangential velocity $\vt=0$) can bear any tangential
    force with absolute value up to $\Ftmax$. If
    sliding occurs ($\vt\neq 0$) the magnitude of the tangential force is
    $\Ftmax$ while its direction is opposite to the sliding velocity.}
   \label{fig:coulomb_graph}
\end{figure}

For the tangential contact force the Coulomb 
friction law has been modified to include the influence of cohesion: If the
relative velocity of the two surfaces is zero (sticking 
contact) the magnitude of the friction force ($\left| \Ft \right|$) can be
any value up to the threshold $\Ftmax$, while in the case of sliding
$\left| \Ft \right| = \Ftmax$, and its direction is opposite to the relative
velocity (\fig{coulomb_graph}). The maximal friction force is proportional
to the normal force including $F_{\rm coh}$:
\begin{eqnarray}
  \label{eq:Ftmax_coh}
  \Ftmax = \mu (\Fn + \Fc) \ ,
\end{eqnarray}
thus the threshold value vanishes when the contact opens ($\Fn = -\Fc$).

\begin{figure}
   \centering
   \includegraphics[scale=0.31]{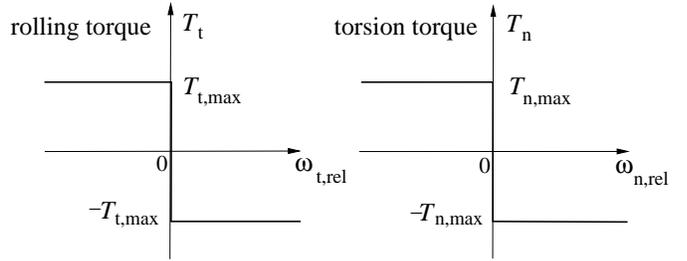}
   \caption{Contact torques applied to suppress rolling and torsion. The
   rules are similar to those of the tangential force. The rolling torque
   (torsion torque) is coupled to the tangential (normal) relative angular
   velocity $\omega_\mathrm{t,rel}$ ($\omega_\mathrm{n,rel}$). For
   definition of the threshold values $\Tnmax$ and $\Ttmax$ see the text.} 
   \label{fig:roll_tors_torque}
\end{figure}

\begin{figure}
\centering 
  \includegraphics[scale=1.3]{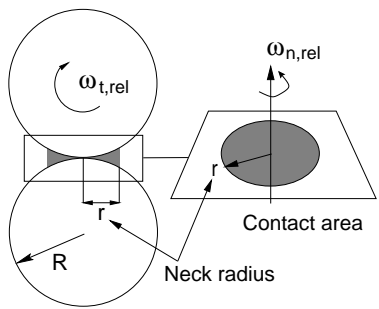}
 \caption{Schematic figure of a sinter neck between two particles.}
 \label{fig:torsion_model}
\end{figure}

The contact laws for the normal and tangential torques are chosen in
analogy to the Coulomb friction law (see \fig{roll_tors_torque}). The
threshold values for the torques $\Tnmax$ and
$\Ttmax$ are defined as force times length, where the force-scale is again
given by
$(\Fn + \Fc)$ and the radius of the sinter-neck determines the relevant
length-scale (Fig. \ref{fig:torsion_model}). This length makes sense only if it is below
the particle size, and in our study we choose it within the range between zero
and the effective radius at the
contact: $\reff= \left( 1/R_1 + 1/R_2\right)^{-1}$, which for two identical spheres
is half the particle radius. The maximal normal and tangential torques
in the model are thus given by:
\begin{eqnarray}
  \label{eq:max_torques}
  \Tnmax &=& \mn (\Fn + \Fc)\reff\\
  \Ttmax &=& \mt (\Fn + \Fc)\reff\ ,
\end{eqnarray}
where the dimensionless parameters $\mn$ and $\mt$ make it possible to
control the strength of blocking, similar to the friction coefficient
$\mu$  in Eq.~(\ref{eq:Ftmax_coh}).

\bigskip

This model is certainly oversimplified, and the physics of
nano-particles might be better described by more sophisticated, yet
unknown microscopic models. Moreover, in our model rolling, torsion
and sliding modes are independent of each other. In general this is
not the case as demonstrated for an ordinary disk on a flat surface 
\cite{farkas2003,Voyenli1985,Goyal1991}, where the coupling between
torsion and sliding
friction reduces both of them. In this case assuming independence
would overestimate friction. If this is generally true, the porosities
obtained in our simulations are expected to be upper bounds.
With some caution we can therefore use our simplified model to analyze
the effect of rolling and torsion friction on the porosity.

\section{Computer Simulations}

The discrete element simulations presented here were performed with
the contact dynamics method \cite{moreau88,moreau94,jean99}, where the
particles are modeled as 
non-deformable spheres and the interaction between particles
is determined by constraint conditions. The application of constraint
forces and torques and an implicit time stepping scheme make this method
especially suitable for the implementation of threshold
dynamics and for the proper treatment of static forces and torques in
blocked or jammed states.

More about our algorithm can be found in
\cite{Unger03a,unger2002,PhDkadau2003}. These works contain the
description of the basic 3D algorithm for ordinary granular materials, an
analysis of the method and its extension to the case of nano-particles.

The system under consideration consists of spherical particles with the same
radius $R$. Starting from a low density state the compactibility  is
measured for various values of $\mt$ and $\mn$, while the coefficient of
friction, $\mu=0.3$, is kept constant. 

The initial configuration is prepared by ballistic deposition
\cite{meakin87,meakin91,Kadau2003}: Particles fall vertically towards
a horizontal plane, one by one with randomly chosen
$x-y$-coordinates. As soon as a falling particle comes closer than a
capture radius $r_\mathrm{capt}$ to the deposit or substrate, 
the contact is established immediately, and the particle sticks
irreversibly. Then the next particle is dropped. Here we choose
$r_\mathrm{capt}=3R$. As a result one obtains deposits
of very low density with tree-like structures.
Using this type of initial configuration is motivated by
filter processes extracting nano-particles out of a gas flow
\cite{lantermann,filippova97}. 

\begin{figure}
\centering
  \includegraphics[width=7cm]{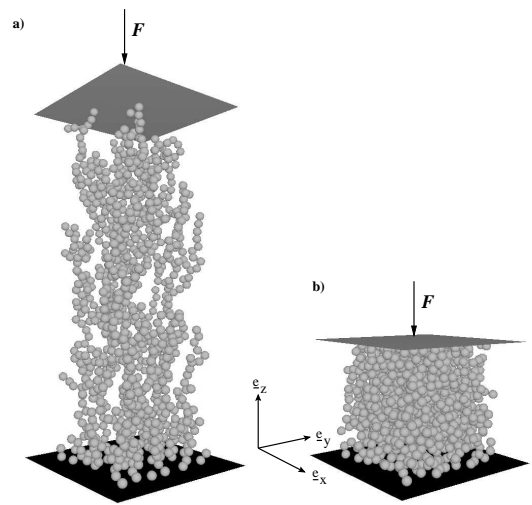}
 \caption{a) Initial arrangement with $1015$ ballistically
 deposited particles. b) Final configuration of the system compressed by
 the external force $F$.} 
\label{fig:3D_compaction}
\end{figure}

We simulate the uniaxial compression of this ballistic deposit by a piston
moving along the $z$-direction towards the bottom plane, with periodic
boundary conditions in $x$- and $y$-direction.  Gravity is neglected. The
compression is caused by a constant pressure $F/L_x L_y$ on the piston
(\fig{3D_compaction}). The simulation ends, when the system reaches a
blocked state, i.e. when the piston comes to rest. The system has size $L_x
= L_y = 25 R$, contains $1015$ particles and the mass of the piston is chosen
to $1000 \rho R^3$, where $\rho$ denotes the mass density of the
particles. The pressure on the piston is $F_{\rm coh}/(400 R^2)$ in all
cases. This is a very weak compacting pressure, as the typical distance
between branches of the ballistic deposit is of the order of $5R$, so that
the typical load on a single branch, $F_{\rm coh} 25/400 = F_{\rm coh}/16$,
is much smaller than the intrinsic force scale given by the cohesion force
$F_{\rm coh}$.

The mesh of values $(\mu_{\rm t}, \mu_{\rm n})$, for which we
determined the final porosity, is indicated in
Fig.\ref{fig:3D_result}. Assuming a smooth dependence of the porosity
on the friction parameters, we can estimate the error bars from
Fig.\ref{fig:3D_result} without the need to do more than one run for
each data point.

\section{Results}

\begin{figure}
\centering
  \includegraphics[scale=0.4]{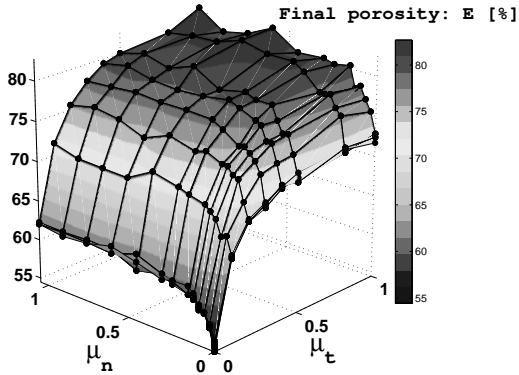}
 \caption{\label{fig:3D_result} Final porosity of the compacted system
   obtained for various 
   values of rolling ($\mt$) and torsion resistance ($\mn$).}
\end{figure}

Based on the final position of the piston we measured the porosity
$\por$ of the blocked states:
\begin{equation}
\por = \frac{V-V_\mathrm{grains}}{V}\ ,
\label{eq:porosity}
\end{equation}
i.e. the relative free-volume in the system (here $V$ denotes the
volume between the bottom and the piston). \fig{3D_result} shows that
contact torques have significant effect on 
the compactification: Whereas without rolling and torsion friction the
final porosity is $\por_0 \approx 54$\% for the weak compaction
pressure we considered, porosities as high as $82$\% are stable
for $\mt = \mn = 1$. 

The porosity added due to rolling and torsion friction,
\begin{equation}
\por_1(\mt,\mn) =\por - \por_0 ,
\end{equation}
saturates in the region where the coefficients $\mt$ and $\mn$ are larger than
$0.4$. More than $\por_1(1,1) \approx 28\%$ cannot be achieved based on the
contact torques, not even if one increases the threshold values far
beyond the physically meaningful range. The maximum porosity is less
than the one of the initial configuration: Suppression of rolling
and torsion degrees of freedom alone does not suffice to avoid
compaction completely.

Qualitatively, rolling friction alone ($\mu_{\rm n} =0$) as well as
torsion friction alone ($\mu_{\rm t} =0$) have similar effects on the
porosity. Quantitatively we find, that $\por_1(1,0) \approx
18\%$ is about twice as big as $\por_1(0,1) \approx 8\%$. The
reason is not clear, but it is intriguing to notice that rolling
friction $\mt$ suppresses two degrees of freedom, while $\mn$
suppresses only one. The ratio
$\por_1(\tilde\mu,0)/\por_1(0,\tilde\mu)\approx 2$ 
is roughly independent of $\tilde\mu$.
 
An interesting property of the
function $\por_1(\mt,\mn)$ is, that it can be very 
well represented by the following sum:
\begin{equation}
  \label{eq:porosity_sum}
  \por_1(\mt,\mn)=\por_1(\mt,0)+\por_1(0,\mn) \  .
\end{equation}
This shows that rolling and torsion friction contribute independently
to the porosity. The difference $|\Delta E|$ between the two sides in
Eq.~\eq{porosity_sum} is less than about $3$\%,
 and $\Delta E$ fluctuates around zero with no apparent systematics.

Finally we would like to discuss, how the static response of the
system to the external load differs with and without contact
torques. In the absence of contact torques, strong tensile contact forces 
appear and seem to be crucial for stabilizing the compressive force
lines against buckling in the porous system. This pore stabilization
mechanism has also been found in two-dimensional systems \cite{Kadau2003}.
Of course, the contribution of the compressive forces to the
macroscopic stress tensor overcompensates the one of the tensile
forces in order to balance the external load (table \ref{tab:stress}).  

\begin{table}
\centering
\begin{tabular}[b]{|cc|ccc|}
\hline
~~ $\mt$ ~~ & ~~ $\mn$ ~~ & ~~ $\sigma_{zz}$ ~~ & ~~ $\sigma_{zz}^+$ ~~ & ~~
 $\sigma_{zz}^-$ ~~ \\
\hline
0.0    &        0.0&            1.0&    2.2&    -1.2\\
0.1    &        0.1&            1.0&    1.4&    -0.4\\
0.3    &        0.3&            1.0&    1.1&    -0.1\\
1.0    &        1.0&            1.0&    1.1&    -0.1\\
\hline
0.0    &        0.3&            1.0&   2.2&   -1.2\\
0.0    &        1.0&            1.0&   1.9&   -0.9\\
\hline
0.3    &        0.0&            1.0&   1.2&   -0.2\\
1.0    &        0.0&            1.0&   1.4&   -0.4\\
\hline
\end{tabular}
\caption{ The influence of the normal and tangential contact torques is
  shown on the stress transmission along the direction of the
  uniaxial compression. $\sigma_{zz}^+$ and $\sigma_{zz}^-$ are 
  containing only contributions of the compressive and the tensile contacts
  respectively.
\label{tab:stress}}
\end{table}

In the case where contact torques are allowed, rolling and torsion
friction already stabilize the force lines against buckling, before
significant tensile forces develop. 
This does not mean that cohesion is superfluous in this
case as it is also responsible for the enhanced threshold values of the
contact torques (Eq.~\eq{max_torques}). 

We measured the macroscopic stress tensor (\cite{Christoffersen81}):
\begin{equation}
\sigma_{ij} = - \frac{1}{V} \sum_{c^{(+)}} F^c_i \, l^c_j - \frac{1}{V}
\sum_{c^{(-)}} F^c_i \, l^c_j = 
\sigma^+_{ij} + \sigma^-_{ij} \, , 
\label{eq:stress}
\end{equation}
separately taking only the contacts under compression
($\sigma_{ij}^+$) or the ones under tension ($\sigma_{ij}^-$) into
account. (Here $l^c_j$ is
the $j$th component of the interstice vector connecting the centers of the
particles at contact $c$.)
The $zz$-components of the stress tensor are shown in the table
(\ref{tab:stress}), where 
the resulting 
stress $\sigma_{zz}= F/L_x L_y$ is of course determined by the
pressure on the piston, which is the same for all cases and is taken
as unit stress. In the zero-torque case more than twice
of the external pressure is provided by the contacts under compression, 
because a sufficient amount of tension must be allowed in the system
in order to stabilize the pores.
This internal counter-stress is diminished significantly by
rolling friction, whereas torsion friction alone has little effect on 
the partial stresses $\sigma_{zz}^+$ and $\sigma_{zz}^-$.

\section{Conclusion}

We presented 3D contact dynamics simulations of cohesive powders and
studied the porosity under weak uniaxial compression. We focused on the
effect of 
contact torques which suppress relative \emph{torsion} and \emph{rolling}
of the adjacent particles. It was found that the presence of contact
torques has enormous impact on reducing the final density of the system and in
addition the characteristics of the stress transmission is altered: Without
torsion and rolling friction strong tensile forces develop and play
important role in the mechanical stabilization. These tensile contacts,
however, cancel a large part (more than half) of the pressure exerted by
compressive forces and only the remaining part is utilized to resolve the
external load. This situation is changed by switching on contact torques,
which reduce tensile forces significantly in the system. In that case the
stress due to the compressive contacts corresponds approximately to the
external load and only a small part of this stress is ``wasted'' to
overcompensate tensile forces.

A remarkable feature is the additivity found in the porosity, i.e.\ the
porosity is well represented as the sum of independent contributions of
the torsion and rolling friction. Our results suggest that each
relative motion mode, if it is 
suppressed, results in an additional free volume in the system
independent of the other modes. Whether this picture holds also for
translational degrees of freedom (i.e.\ altering the Coulomb friction
coefficient or the cohesion force) is a subject of further investigation.

\bibliographystyle{unsrt}
\bibliography{paper_tors}

\end{document}